\documentclass[aip,jcp,reprint]{revtex4-1}

\draft 

\usepackage{natbib}

\usepackage{graphicx}
\usepackage{dcolumn}
\usepackage{bm}

\usepackage{amsmath}
\usepackage{bm}
\usepackage{chemarr}
\usepackage[usenames,dvipsnames]{color}

\usepackage{graphicx}
\usepackage{subfig}
\usepackage{epsf} 

\graphicspath{ {./images/} {./} }

\begin{document}


\title{Positive feedback can lead to dynamic nanometer-scale clustering on cell membranes}



\author{Martijn Wehrens}
\affiliation{FOM Institute AMOLF, Science Park 104, 1098 XG Amsterdam, The Netherlands}

\author{Pieter Rein ten Wolde}
\affiliation{FOM Institute AMOLF, Science Park 104, 1098 XG Amsterdam, The Netherlands}

\author{Andrew Mugler}
\email[]{andrew.mugler@emory.edu}
\altaffiliation{Present address: Department of Physics, Emory University, 400 Dowman Drive, Atlanta, GA 30322, USA}
\affiliation{FOM Institute AMOLF, Science Park 104, 1098 XG Amsterdam, The Netherlands}



\begin{abstract}
Clustering of molecules on biological membranes is a widely observed phenomenon. In some
cases, such as the clustering of Ras proteins on the membranes of mammalian cells, proper cell
signaling is critically dependent on the maintenance of these clusters. Yet, the mechanism by which
clusters form and are maintained in these systems remains unclear. Recently, it has been discovered that activated Ras promotes further Ras activation.  Here we show using particle-based simulation that this positive feedback is sufficient to produce persistent clusters of active Ras molecules at the nanometer
scale via a dynamic nucleation mechanism. Furthermore, we find that our cluster statistics are consistent with experimental observations of the Ras system. Interestingly, we show that our model does not support a Turing regime of macroscopic reaction-diffusion patterning, and therefore that the clustering we observe is a purely stochastic effect, arising from the coupling of positive feedback with the discrete nature of individual molecules. These results underscore the importance of stochastic and dynamic properties of reaction diffusion systems for biological behavior.
\end{abstract}


\maketitle 

%


\section{Introduction}

%

%
%

Clustering of molecules on biological membranes is a widely observed phenomenon
which is known to be important for signaling \cite{Kholodenko2010, Grecco2011, Sourjik2010, Mugler2013r}.
Many mechanisms are implicated
in cluster formation, including membrane rafts \cite{Pike2009, Lingwood2010}, interactions with
the cytoskeleton \cite{Kusumi2005, Plowman2005, Machta2011}, complex formation \cite{Murakoshi2004}, and binding to scaffold
proteins \cite{Dard2006}.
%
%
More recently, the reaction-diffusion dynamics of the underlying biochemical network 
have begun to be investigated as another possible mechanism behind membrane clustering.
In particular, it has become appreciated that the clustered component is 
often subject to positive regulatory feedback \cite{Das2009, Jilkine2011}.
%
%
This raises the interesting possibility
that clustering is not only imposed on particles by
rafts, the cytoskeleton, or scaffolds,
but can also arise intrinsically from, or be further amplified by, the reaction-diffusion dynamics of the feedback network.
%
%
Positive feedback
has long been known to 
be critical for
pattern formation in macroscopic systems, where the large numbers of particles can be
approximated as a continuum \cite{Murray2003a}.
In contrast, however, particle numbers in clusters on biological membranes can be small, even as small as a few to tens of molecules.
This raises the question whether positive feedback can lead to clustering in such systems, where noise from diffusion and reactions dominate the dynamics \cite{Das2009, Jilkine2011}.

One of the model systems in which positive feedback was discovered to play an important role is the Rat sarcoma
(Ras) signaling system.
The Ras protein is found in small nanoclusters on the cell membrane 
which are tens of nanometers in size 
and contain a few to tens of molecules, depending on the Ras isoform and the conditions \cite{Eisenberg2011a, Tian2007, Plowman2005, Prior2003}.
%
%
%
Ras, a GTPase, performs its signaling function by switching between
its inactive GDP-bound state and its active GTP-bound state. 
Activation is catalyzed by a class of proteins called Guanine Exchange Factors (GEFs),
of which a well-studied example is Son of Sevenless (SOS).
It was recently discovered that SOS has two binding sites for Ras \cite{Margarit2003, Groves2010}. 
One is a catalytic binding site which performs the 
activation.
%
The second is an allosteric binding site that can bind Ras-GTP.
This binding increases 
the catalytic activity of the activation site,
%
thus introducing 
a positive feedback loop: the presence of active Ras-GTP
will increase the (local) production of Ras-GTP.


It has been suggested
that the Ras positive feedback mechanism can
cause clustering of active Ras particles. 
%
Simulations by Das et al.\ show that the positive feedback loop in the biochemical network
of the 
Ras protein
leads to the growth of domains of active molecules,
even when the Ras-GDP and Ras-GTP have the same diffusion constant \cite{Das2009}.  
However, the behavior of this system in steady state was not investigated (deactivation was not modeled),
and domain size was not compared with experiments.
Other recent simulations by Jilkine et al.\ show that a positive feedback loop in the biochemical network
can lead to 
clustering,
given that the diffusion constants of the Ras-GDP and Ras-GTP species are unequal
\cite{Jilkine2011}.
Their study demonstrates clustering under steady-state conditions, but does not 
quantify the clustering or compare it with experimental values.
An important open question remains whether a positive feedback loop in the Ras reaction-diffusion network
can generate clusters
that shows the same
statistics as are observed in biological systems.
To investigate the effects of positive feedback, we introduce a model of the Ras biochemical network containing the minimum number of chemical reactions required to incorporate the observed positive feedback loop.
%
%
We employ the distribution of interparticle distances, 
as well as a measure of clustering based upon the work of Hackett-Jones et al.\ \cite{Hackett-Jones2012}, 
to quantify clustering of particles in our simulated system when it is in steady state.
Importantly, we show that the positive feedback loop found in the Ras system can indeed produce clusters that have the same statistics as observed in experiments.

Our simulations reveal clustering in a biologically relevant range of ratios between the Ras-GDP and Ras-GTP diffusion constants.
Clustering is observed when Ras-GTP has a ten times slower diffusion constant than Ras-GDP,
but importantly also when the Ras-GTP diffusion constant is equal to the Ras-GDP diffusion constant.
%
%
The latter is important because
it is known that reaction-diffusion systems can exhibit clustering when two species have
different diffusion constants \cite{Murray2003a},
but 
it remained unclear whether clustering
can be observed
in steady state for species with the same diffusion constant.
%
Concerning the Ras system, single molecule tracking experiments suggest that Ras-GTP molecules slow down compared to Ras-GDP molecules:
they either diffuse 3-4 times slower than Ras-GDP or become immobile entirely \cite{Murakoshi2004, Lommerse2005}.
Membrane domains, scaffold proteins and actin filaments are implicated in this mechanism \cite{Murakoshi2004, Lommerse2005, Rotblat2004}.
Diffusion dynamics might however be complex.
It is argued that Ras-GDP and Ras-GTP both have a slow and fast moving fraction of molecules \cite{Lommerse2005}.
This raises the question of how robust the observed clustering is to the precise diffusion dynamics.
Our simulations reveal clustering for a range of ratios of diffusion constants,
indicating that the positive feedback clustering mechanism we study here is very resilient to changes in diffusion dynamics.

Finally,
we show that the clustering we observe is not predicted by macroscopic theory.
While our minimal model contains qualitative features associated with the macroscopic Turing clustering mechanism (local positive feedback and species-dependent diffusion constants), 
we show analytically that a macroscopic description of our reaction-diffusion system does not have a Turing clustering regime.
Previous work has shown that fluctuations inherent to discrete particle systems can extend the regime in which clustering is observed beyond the macroscopic Turing regime associated with the model \cite{Butler2009, Butler2011}.  Here we show that clustering can emerge from such fluctuations in the absence of a Turing regime altogether.
%
We thus reveal a surprising dynamic mechanism of clustering,
which is entirely due to stochastic effects,
and which results in steady-state clusters
with the same statistics 
as observed in experiments.

\section{Model}


%

We consider a model of Ras activation which is minimal, but nonetheless captures the positive feedback resulting from interaction with the activator SOS, and moreover whose parameters remain directly informed by experimental measurements.
 
\begin{figure}
\begin{center}
  \includegraphics[width=0.45\textwidth]{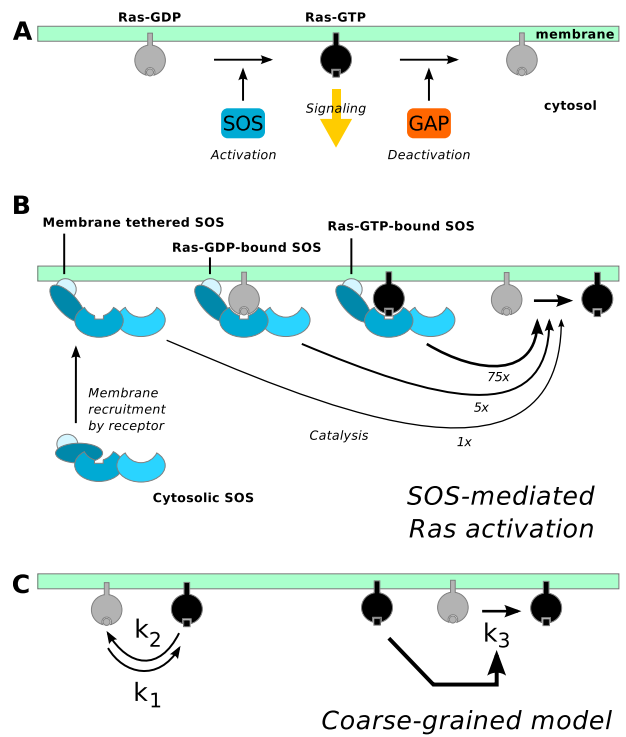}
\end{center} 
\caption{
%
%
Schematic of Ras activation and inactivation.
\textbf{A.} 
Ras is activated by guanine exchange factors including SOS and 
deactivated by GTPase activating proteins (GAPs).
%
\textbf{B.} 
Allosteric binding of Ras-GDP or Ras-GTP to SOS increases its rate of further Ras activation by roughly $5$- and $75$-fold, respectively; the latter effect introduces the positive feedback.
%
\textbf{C.} 
%
Our minimal model coarse-grains over the SOS and GAP, while retaining the positive feedback, as described in the text.
%
%
}
\label{fig:SOSRas}
\end{figure}

The mechanism by which Ras is believed to be activated is shown in Fig.\ \ref{fig:SOSRas}A.
Ras is membrane-bound and exists in either a GDP-bound (inactive) or a GTP-bound (active) state. 
Activation occurs when 
%
stimulated receptors
recruit the 
activator SOS to the membrane (Fig.\ \ref{fig:SOSRas}B). 
SOS activates Ras by catalyzing the release of GDP from Ras.
Since GTP is present in an about tenfold higher concentration, 
Ras will then subsequently bind GTP and become active \cite{Lodish2000}.
Importantly, in addition to this catalytic domain, it was recently discovered that SOS contains an additional allosteric binding pocket for Ras \cite{Margarit2003, Groves2010}.
This pocket can bind either Ras-GDP, resulting in 5-fold higher catalytic activity than when unbound,
or Ras-GTP, resulting in 75-fold higher catalytic activity (Fig.\ \ref{fig:SOSRas}B) \cite{Sondermann2004, Freedman2006}.  This latter reaction introduces the positive feedback loop: the more Ras-GTP in a local area, the faster it is produced by Ras-GTP-bound SOS.
%
%
%
%
Ras has an intrinsic GTPase activity, and Ras-GTP is thus always slowly converted to Ras-GDP.  GTPase-activating proteins (GAPs) deactivate Ras by greatly increasing the rate of this reaction (Fig.\ \ref{fig:SOSRas}A).
%
%
%
%
More detailed aspects of the Ras system, such as
the presence of activators other than SOS \cite{Freedman2006, Das2009a} and the fact that Ras exists in three isoforms that differ in their localization \cite{Eisenberg2011a, Prior2003, Plowman2005}, are omitted from this minimal model.

Striving for simplicity, we consider a model which coarse-grains out the SOS and GAP, but retains the positive feedback, as shown in Fig.\ \ref{fig:SOSRas}C.  The model contains only two species, Ras-GDP (D) and Ras-GTP (T):
%
\begin{equation}\label{eq:eqreaction}
  D \xrightleftharpoons[k_2]{k_1} T 
,
\qquad
  D + T \xrightarrow{k_3} T + T 
.
\end{equation}
Deactivation, at rate $k_2$, is spontaneous and independent of space, which is valid for a GAP which is fast-diffusing in the cytoplasm, or at sufficiently low GAP concentrations that the intrinsic Ras GTPase activity dominates.  Activation by SOS not bound to Ras occurs at rate $k_1$, while activation by Ras-GTP-bound SOS occurs at the faster rate $k_3$.  The latter reaction requires interaction of a D and a T molecule, and thus introduces the space-dependent positive feedback.  Coarse-graining over SOS, as done previously \cite{Jilkine2011}, is valid when (i) the abundance of SOS is large, and/or (ii) the diffusion of SOS is fast, such that SOS binds D or T on a timescale faster than those of the activation reactions.  As also done previously \cite{Das2009}, we ignore activation by Ras-GDP-bound SOS, since (i) Ras-GDP is ten times less likely to bind to the allosteric site of SOS than Ras-GTP \cite{Sondermann2004}, and (ii) activation by Ras-GDP-bound SOS is only 5 times faster than activation by 
unbound SOS, 
whereas activation by Ras-GTP-bound SOS is 75 times faster \cite{Sondermann2004, Freedman2006}.


Experimental measurements constrain the model parameters.
The radius of both D and T is set by the measured radius of gyration of Ras, $1.7$ nm \cite{Fujisawa1994}.
The density of Ras on the membrane is set by observations that Ras occupies a surface fraction of roughly $1\%$ \cite{Tian2007}.
The diffusion constant of Ras has been measured to lie between $0.01$ $\mu$m$^2$/s and $1$ $\mu$m$^2$/s \cite{Das2009}; therefore we set the diffusion constant of D molecules to $0.1$ $\mu$m$^2$/s.
There is evidence that the diffusion of Ras slows down upon activation \cite{Murakoshi2004, Lommerse2005}; therefore, the ratio of diffusion constants of T to D is varied from $0.1$ to $1$.
The deactivation rate $k_2$ is set to yield an active fraction of $[T]/([T]+[D])\sim0.10-0.15$ in steady state, which is consistent with a typical experimentally observed fraction of $\sim$$20\%$ \cite{Hancock2005}.
The rate $k_3$ is set such that
the bimolecular reaction is placed in the diffusion-limited regime, since this is the regime in which we expect to observe feedback-induced clustering.
The measured 75-fold speedup of activation \cite{Freedman2006} sets the ratio $k_3[T]/k_1\sim 75$, which together with $k_3$ and $[T]$, determines the basal activation rate $k_1$.

\section{Methods}


We simulate our model system in two dimensions using the enhanced Greens Function Reaction Dynamics (eGFRD) scheme \cite{Takahashi2010, Sokolowski2013, Sokolowski2013a}.
Clustering naturally induces high local molecular densities, for which we revert to the
Brownian Dynamics (BD) algorithm \cite{Paijmans2012} available within the eGFRD framework.
%
%
%
%
%
%
A total of $N=125$ spherical particles of radius $a = 1.7$ nm are simulated with periodic boundary conditions in a square area of side length $L = 337$ nm, to give the observed surface fraction of $N(\pi r^2)/L^2 = 0.01$.
We run $17$ simulations, each of which are initialized with $(N-1)$ D particles and $1$ T particle, which are placed at random in the simulation plane.  Each simulation is run for $0.13$ seconds of simulated time to reach steady state, and then for an additional $5.07$ seconds ($975$ timepoints).  

Parameter values, constrained by experiments as described above, are as follows.  The diffusion constant of D molecules is $\kappa_D=0.1$ $\mu$m$^2$/s, and the diffusion constant of T molecules is $\kappa_T = R\kappa_D$, where the ratio $R$ is varied as $R=\{0.1, 0.25, 0.5, 0.75, 1.0\}$.  The reaction rates are $k_1=18.21$ s$^{-1}$, $k_2 = 770.87$ s$^{-1}$, and $k_3 = 10$ $\mu$m$^2$/s.  These values result in the ratios $\kappa_D/k_3 = 0.01 \ll 1$, which places the system in the diffusion limited regime, and $k_3[T]/k_1 = 82.2$ (where $[T]$ is the steady-state averaged concentration), which is close to the measured value of $75$.
%
%

Since eGFRD does not accommodate two reactants forming two products, the $k_3$ reaction in Eq.\ \ref{eq:eqreaction} is split into two reactions: one for complex formation and one for the reaction resulting in the product, with respective rates $k_{3a}=k_3$ and $k_{3b} = 10^9 \times k_3 N/(2L^2)$. 
This latter choice makes the $k_{3b}$ reaction faster than all other timescales and thus ensures that the $k_{3a}$ reaction is the rate-determining step.

For simulations without positive feedback ($k_3=0$), the reaction rates are adjusted, such that the steady-state $[T ]$ remains the same as with positive feedback. 
Specifically, $k_1\to k_1^- = k_1 + k_3 [T]$,
in order to keep the same rate of Ras activation,  and
$k_2\to k_2^-$, where $k_2^-$ is determined by the mean-field steady-state condition
$[T] = (N/L^2) k_1^- / (k_1^- + k_2^-)$. Thus,
$k_1^-$ is set to $1604$ s$^{-1}$ and $k_2^-$ is set to $9536$ s$^{-1}$ for these simulations. We run 12 simulations without positive feedback.

\section{Results}



\subsection{Pair-wise distance distribution reveals clustering}
\label{sec:Pr}

%
To characterize clustering, we look at the distribution of pair-wise distances $P(r)$.
We note that when normalized by the distance distribution for a set of randomly positioned particles (an ``ideal gas''), $P(r)$ becomes the pair correlation function $g(r)$, also called the radial distribution function.  The pair correlation function is a staple of statistical mechanics, often used to understand the packing properties of gases, liquids, solids, and other many-particle systems.
 
%
%
%
%
%

\begin{figure*}
\begin{center}
\subfloat{\label{fig:Pr1}\includegraphics[width=0.90\textwidth]{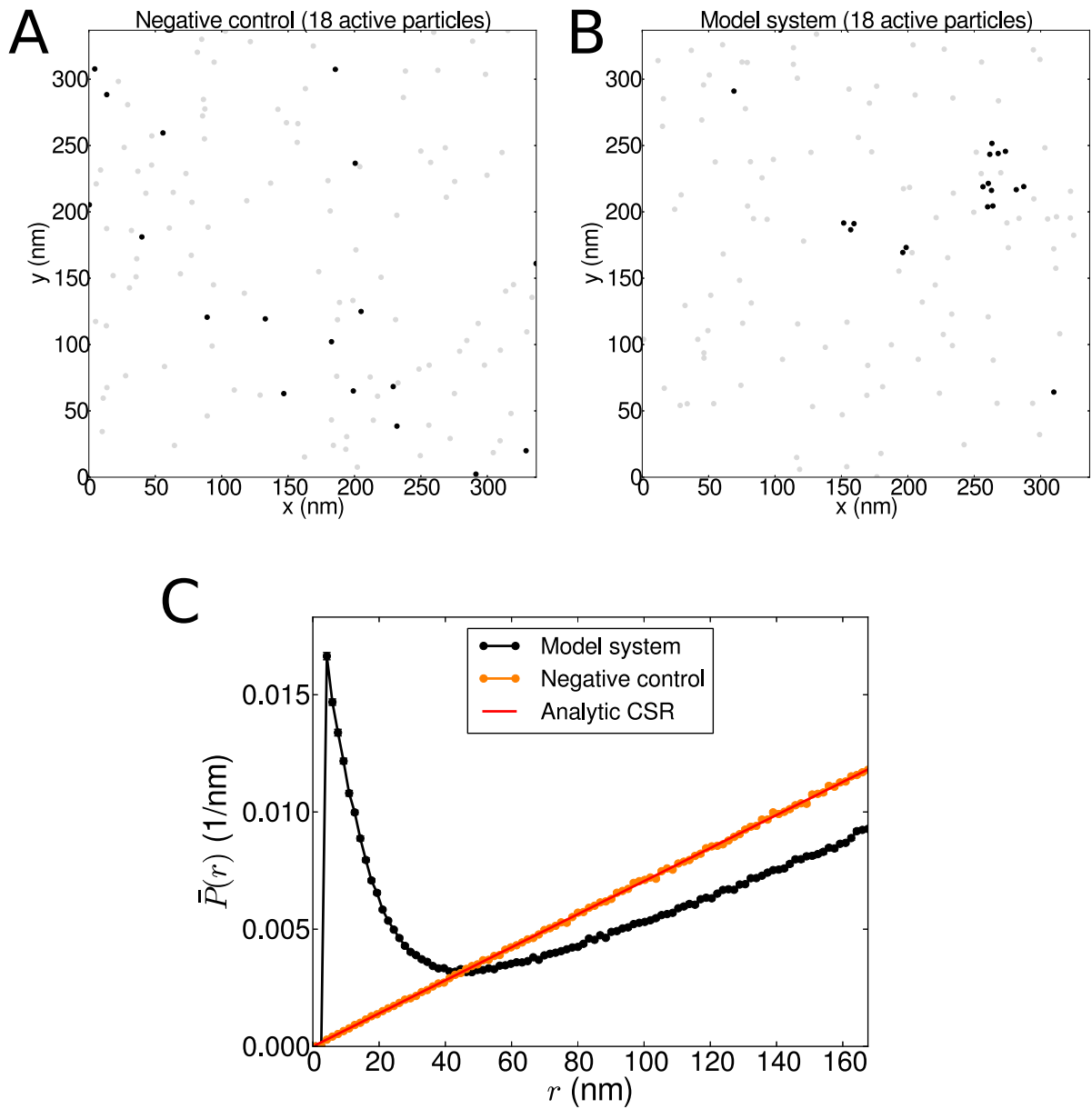}}
%
\end{center} 
\caption{
Positive feedback results in persistent clustering in steady state.
\textbf{A.}
A snapshot from a simulation without feedback (negative control) illustrates a random distribution of active Ras molecules (black circles).
\textbf{B.}
In contrast, a snapshot from a simulation of the model system, with feedback, illustrates small clusters of active Ras molecules.
\textbf{C.} 
The distribution of interparticle distances $\bar{P}(r)$ for the negative control agrees with the complete spatial randomness (CSR) analytic expectation (Eq.\ \ref{eq:PCSR}), whereas $\bar{P}(r)$ for the model system deviates sharply from the CSR expectation due to the clustering.  Parameters are as in {\em Methods}, with $R=0.1$.  In A and B, black and grey circles are active and inactive Ras molecules, respectively.  In C, $\bar{P}(r)$ is averaged over all $975$ snapshots, and error bars are standard error of the mean determined from 17 independent simulations.
}
\label{fig:Pr}
\end{figure*}

Figure \ref{fig:Pr}A and B show snapshots of simulations in the absence and in the presence of positive feedback, respectively.  We refer to the case without feedback as the negative control.  Qualitatively, it is already apparent from Fig.\ \ref{fig:Pr}A and B that, at the particular moment in time shown, small groups of active molecules are closer together in the presence of feedback than in the absence of feedback.  Mechanistically, this is because a single active molecule was activated spontaneously, and the positive feedback caused neighboring molecules with which it interacted to become activated at a faster rate, thus nucleating a cluster.  To be sure that this apparent clustering behavior is significant and persists in steady state, and moreover to compare with the statistics of experimental data, we turn to $P(r)$.  
%
%
%
%
%

For all particles in a given snapshot we compute $P(r)$, which, upon discretizing in bins of width $\Delta r$, reads
\begin{equation}
\label{eq:Pr}
 P(r)\Delta r \equiv  \frac{1}{N(N-1)} \sum_{i=1}^{N} m_i(r) \Delta r 
.
\end{equation}
Here $m_i(r)\Delta r$ is the number of particles between distance $r-\Delta r/2$ and $r+\Delta r/2$
from particle $i$, and $N$ is the number of particles.
The factor $N(N-1)$ normalizes for the $N$ comparisons of each particle with its $N-1$ neighbors.  We compute the average distribution $\bar P(r)$ over all $975$ snapshots of a given simulation in steady state.  We then repeat for $17$ simulations to obtain a standard error for $\bar P(r)$.  Further details on the averaging procedure are given in Appendix \ref{app:averaging}.

For a complete spatial random (CSR) distribution in two dimensions, $P(r)$ is simple and known: the probability to find a particle 
scales with
the area $2\pi r\Delta r$ located in the annular domain between $r-\Delta r/2$ and $r+\Delta r/2$,
%
\begin{equation}
\label{eq:PCSR}
 P_\text{CSR}(r)\Delta r = \frac{2\pi r\Delta r}{L^2}
 ,
\end{equation}
%
where the division by $L^2$ normalizes for the fact that the particles are contained 
on the square simulation plane with both length and width $L$.
We note that particles which have only excluded-volume interactions, but otherwise do not interact with each other, may exhibit deviations from a random distribution. However, this effect manifests itself at higher densities than those considered here. 
The deviations from randomness we observe here are thus induced by the spatio-temporal dynamics of the chemical interactions between molecules.
%
%
%
%
Figure \ref{fig:Pr}C shows $\bar P(r)$, both with feedback and for the negative control, as well as the CSR expression in Eq.\ \ref{eq:PCSR}.  As expected, the negative control tightly follows the CSR curve.  This is because without feedback, there is no interaction between active molecules.  They simply diffuse as hard spheres, and therefore converge statistically to complete spatial randomness.

On the other hand, in the presence of feedback $\bar P(r)$ departs strongly from the CSR curve.  Indeed, $\bar P(r)$ shows a pronounced peak at low distances of roughly $10$ nm.  This is a clear signature of clustering, as it means that particles are significantly more likely to be found at short distances from each other than expected in a random configuration.  We also see that $\bar P(r)$ drops below the CSR curve at large distances.  This is an inevitable consequence of the fact that $\bar P(r)$ is normalized, and it reflects the fact that when molecules are clustered, more small separations must also imply fewer large separations.
Figure \ref{fig:Pr}C demonstrates that our model system produces persistent clustering in steady state.

\begin{figure*}
\begin{center}
  \subfloat{\label{fig:Hrplot1}\includegraphics[width=0.9\textwidth]{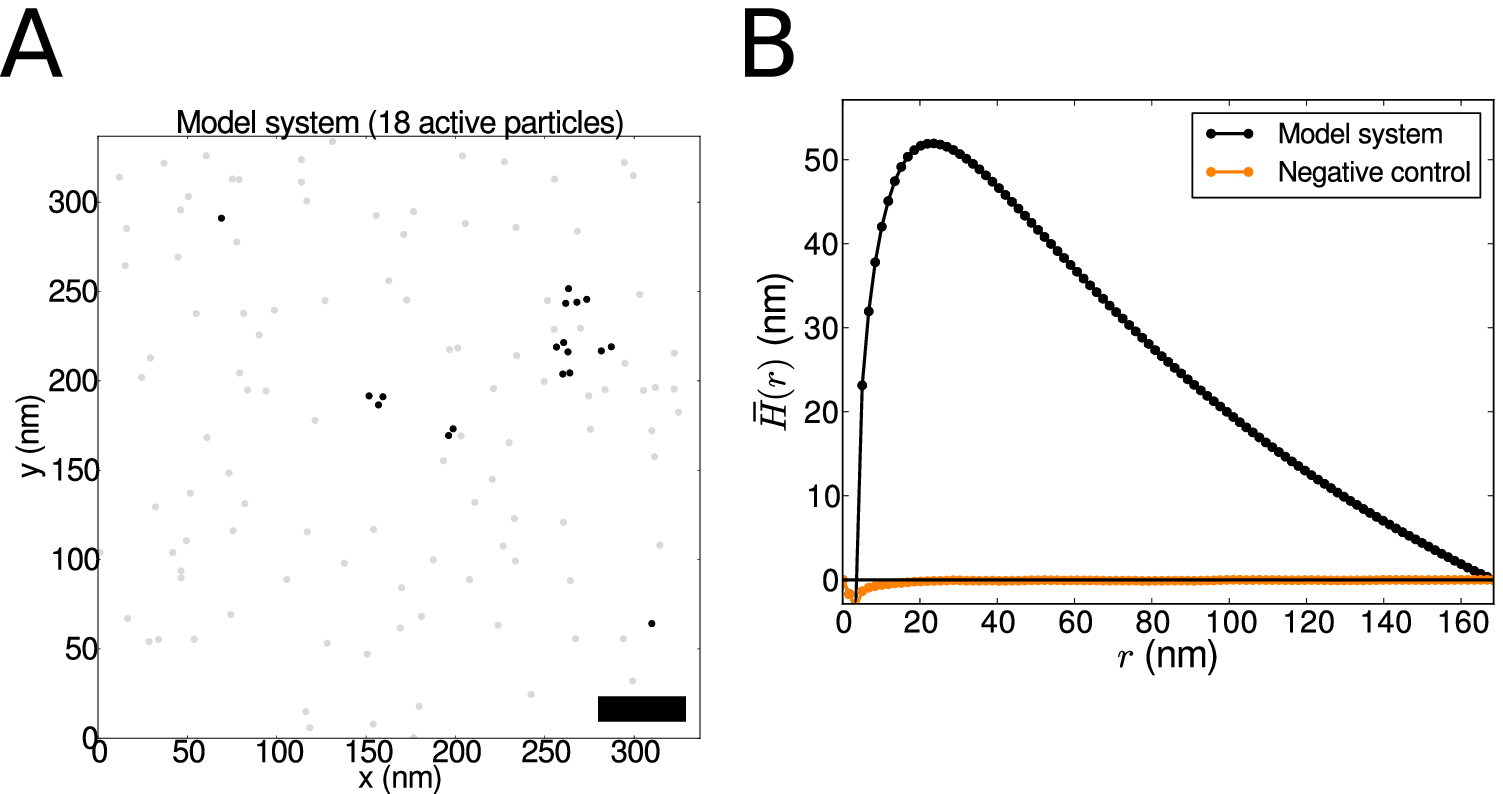}}
    
\end{center} 
\caption{
Clustering statistics from the model are similar to those from experiments.
\textbf{A.} Snapshot from the simulations (identical to Fig.\ \ref{fig:Pr}B).
%
%
\textbf{B.} $H(r)$ function computed from the simulations. Error bars are standard error of the mean, but small and therefore invisible.
%
%
In A and B, parameters are as in {\em Methods}, with $R=0.1$.  In B, $H(r)$ is averaged over all $975$ snapshots.
}
\label{fig:Hr}
\end{figure*}

\subsection{Clustering statistics are consistent with experiments}
\label{sec:Hr}
%
%


We now ask how our simulations compare with experiments on the Ras system. 
Importantly, experiments that study the distribution of active Ras-GTP molecules employ a Ras mutant that is constitutively GTP loaded \cite{Prior2003, Plowman2005, Tian2007}. (It is currently not possible to obtain many-molecule spatially resolved images of only the active form of wild-type Ras.) 
Because it is constitutively active, this mutant should be unaffected by the SOS-dependent switching mechanism that is at the heart of the positive feedback clustering mechanism we study here.
%
Thus, the clustering mechanism studied here cannot be detected in current experimental studies.
Conversely, the experimentally observed clustering cannot be explained by our mechanism.
However, if our simulations of the positive feedback clustering mechanism show similar statistics 
as 
observed in experiments, 
positive feedback clustering might in fact 
significantly contribute to or enhance clustering 
caused by other cell-specific factors.
%

%
Figure \ref{fig:Hr}A shows a snapshot of the simulation (identical to Fig.\ \ref{fig:Pr}B).
%
%
Comparing this figure to experimental data shown in Fig. 1C from Eisenberg et al.\cite{Eisenberg2011a} (top panel), we see a qualitative agreement.
%
However, we
again seek to make a qualitative observation more quantitive using pair-wise statistics.

In the experimental biology literature, clustering is often investigated using a transformed version of $P(r)$ termed Ripley's K-function.  Ripley's K-function is equivalent to the cumulative distribution of inter-particle distances, $K(r) = \int_0^r P(r')dr'$.  Thus, whereas $P(r)dr$ is the probability of a distance lying between $r$ and $r+dr$, $K(r)$ is the probability of a distance lying between $0$ and $r$.  Ripley's K function is widely used in the ecology literature to characterize spatial patterns \cite{Wiegand2004}.

From Eq.\ \ref{eq:PCSR} it is clear that the cumulative function for a CSR distribution is $K(r) = \int_0^r dr'\, 2\pi r'/L^2 = \pi r^2/L^2$.  For this reason, many studies focus on the rescaled quantity $\sqrt{L^2 K(r)/\pi}$, and in particular its deviation from the CSR expectation, $r$.  This deviation is termed the H-function \cite{Kiskowski2009}, and reads
%
%
%
%
%
%
\begin{equation}
\label{eq:Hr}
 H(r) \equiv \sqrt{ \frac{L^2}{\pi} \int_0 ^r P(r') dr'} - r 
.
\end{equation}
A nonzero value of $H(r)$ reflects a deviation from complete spatial randomness at the particular inter-particle distance $r$.
%

Figure \ref{fig:Hr}B shows $H(r)$ for our simulations.
This $H(r)$ function can be compared with 
$H(r)$ functions calculated from experimental data,
such as the one shown in Fig. 1D in Eisenberg et al.\cite{Eisenberg2011a} (the line with filled squares in this plot is the relevant one, as it is based on artificially activated Ras particles; note that $H(r) = L(r)-r$). 
We see that not only are the two plots similar in shape, but both reach a peak around $20$$-$$30$ $\mu$m, which is related to the characteristic cluster size \cite{Kiskowski2009}.  
(The heights of Fig.\ \ref{fig:Hr}B from this manuscript and Fig. 1D from Eisenberg et al. cannot be compared, since the latter is normalized by a $99\%$ confidence value, which is particle number-dependent \cite{Eisenberg2011a, Prior2003}.)
The agreement is particularly noteworthy because the parameters of our model have been set by experimental data from the Ras system where such data are known (see {\it Model} and {\it Methods}).
As mentioned, this agreement indicates that 
positive feedback clustering might in fact 
significantly contribute to or enhance clustering 
caused by other cell-specific factors, a point that is further explored in the discussion.

%
%
%

\subsection{Extent of clustering is consistent with biologically relevant regime}
%
%
%
%
%
%
Figure \ref{fig:Pr}C clearly demonstrates that the distribution of active molecules under our model is different from random, 
but this leaves two open questions: (1) can we quantify this difference and (2) how can we relate this difference to particle configurations? 
To address the first question, we adopt a summary statistic well studied by Hackett-Jones et al.\ \cite{Hackett-Jones2012}.  This statistic quantifies the amount of deviation of the distribution from CSR using a sum of squared differences,
\begin{eqnarray}\label{eq:sigmasquared}
 \bar{\sigma} ^2  
	  & = & \frac{1}{N M} \sum_{i=1}^{N} \sum_{j=1}^{M} {[m_i(r_j)\Delta r - m_\text{CSR}(r_j)\Delta r]^2} 
,
\end{eqnarray}
%
where the first sum averages over the different particles $i$, 
and the second sum runs over the $M$ bins of width $\Delta r$ in which the simulated space has been discretized.
Importantly, due to finite-number noise, even a set of randomly placed particles has a distribution that will deviate from the infinite-particle prediction (Eq.\ \ref{eq:PCSR}), leading to a value of $\bar{\sigma}^2$ that is larger than zero.
%
In fact, Hackett-Jones et al.\ derived an analytical expression for this
value as a function of particle number $N$,
%
%
\begin{equation}
\label{eq:sigmaCSR}
 \sigma ^2 _{\text{noise}} =
\left[
  (N-1) - \frac{(N-1)^2 s}{A} 
\right]
\left[
\frac{1}{M} \sum_{j=1}^{M} {
     \frac{S_j}{A}
     \left(1- \frac{S_j}{A}\right)
 } 
\right]
,
\end{equation}
%
%
with $A=\pi(L/2)^2$ the area of the circle defined by the largest bin,
$s = \pi a^2$ the cross-sectional area of a single particle
and $S_j = 2\pi r_j\Delta r$ the area of the $j$th annular bin at radius $r_j = (j-1/2)\Delta r$.
The right hand bracketed term 
normalizes for bin geometries,
and 
the left hand bracketed term estimates the variance
by assuming the particle count in each bin follows a P\'{o}lya distribution \cite{Hackett-Jones2012}.
%
%
We thus define a normalized index of clustering as
\begin{equation}
\label{eq:chi2}
\chi^2 \equiv \bar{\sigma} ^2 / \sigma ^2 _\text{noise}
,
\end{equation}
which corrects for finite number effects.
%
A $\chi^2$ value of 1 indicates no deviation from CSR, and thus no clustering, whereas a $\chi^2$ value above 1 indicates deviation from CSR, and is consistent with clustering.

When we compute the $\chi^2$ value for our negative control (no feedback), we indeed find that $\chi^2 = 0.9997 \pm 0.0002$ ($99\%$ confidence interval), which is approximately one. This confirms that all deviation from uniformity in the negative control is due to finite-number noise. On the other hand, the value for the model system (with positive feedback and $R = 0.1$) lies at $\chi^2 = 1.167 \pm 0.004$ ($99\%$ confidence interval), significantly higher than 1. This confirms that the configurations observed in our simulation do not follow a CSR distribution, but are instead consistent with clustering. 

\begin{figure*}
\begin{center}
  
  \subfloat{\includegraphics[width=1.0\textwidth]{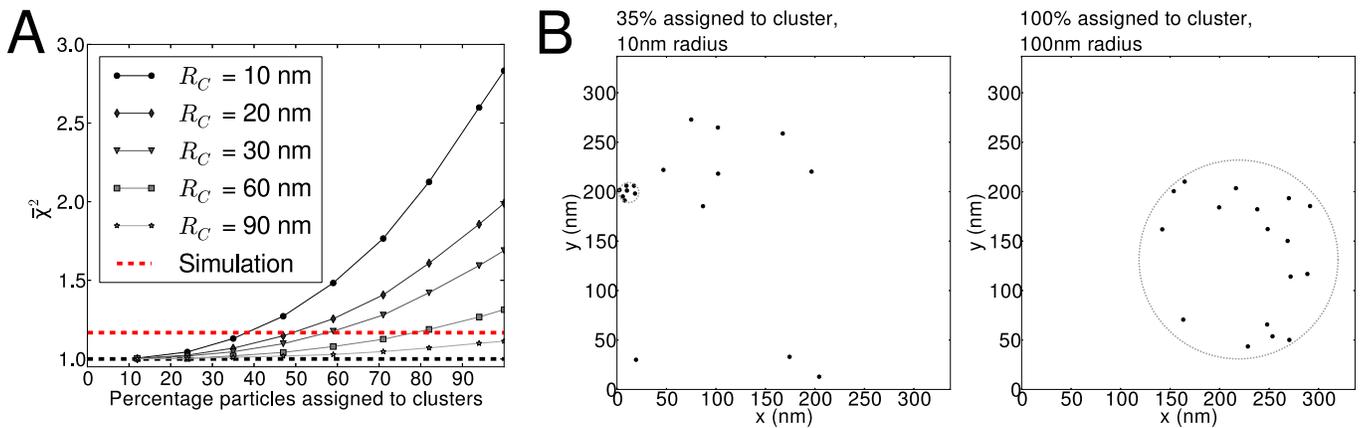}}
\end{center} 
\caption{
Values of clustering index $\chi^2$ were calculated for artificial computer generated configurations that are consistent with experimental observations (see text for more information).
\textbf{A.} 
In black, values of $\chi^2$ are shown for different cluster radii $R_C$ and different percentages of particles participating in the cluster.
In red, the $\chi^2$ value found by simulating the model system, with a ratio of $0.1$ between the diffusion constants, is shown.
The black dashed line lies at $\chi^2 = 1$, the expected value for a random distribution.
\textbf{B.} Two example configurations.
Average $\chi^2$ values have been
determined by averaging over 1000 configurations per datapoint. 
For each configuration, 17 particles were either placed randomly within an area of size $L^2$, or randomly
placed inside the cluster, which was given a random location.  
%
}
\label{fig:figX_phaseplot}
\end{figure*}

To address the second question, namely to what configurations different values of $\chi^2$ correspond, we calculate the value of $\chi^2$ for biologically relevant configurations of clustering. Experiments suggest that $40\%$ of active particles are found in clusters \cite{Tian2007, Plowman2005, Hancock2005},
clusters consist of 6-7 particles \cite{Prior2012, Hancock2005, Plowman2005}, 
and that the radii of these clusters lie between $5-12$ nm \cite{Hancock2005,Plowman2005,Prior2003,Omerovic2009},
although radii of hundreds of nanometers have also been reported \cite{Pezzarossa2012}.
%
We thus randomly generate artificial configurations, consistent with the active particle density in our simulations, with between $10\%$ and $100\%$ of particles placed in a cluster that is between $10$ nm and $100$ nm in radius.
The results are shown in Fig. \ref{fig:figX_phaseplot}A, with two example configurations shown in Fig. \ref{fig:figX_phaseplot}B.
We find that the value of $\chi^2$ observed in our simulations is indeed consistent with configurations suggested by experiments, in which roughly 40\% of particles are confined to a cluster with radius of about $10$ nm (Fig. \ref{fig:figX_phaseplot}A).
%
%
%
%
%
%

%

%
%
%
%

Figure \ref{fig:figX_phaseplot} also demonstrates that many different clustering configurations can share the same value of $\chi^2$.  This point is clear from the dashed red line in Fig. \ref{fig:figX_phaseplot}A, which shows that the value of $\chi^2$ observed in simulations is consistent with a broad class of configurations, ranging from those with a low fraction of particles in a small cluster, to those with a high fraction of particles in a large cluster. 
Indeed, in our dynamic clustering mechanism, there is in fact no well-defined clustered fraction or cluster size.
The observed non-random distribution of particles clearly indicates clustering,
but
every particle has a propensity to participate in clustering, 
and
clusters have no rigidly bounded size.

\subsection{Clustering persists with equal diffusion coefficients}

\begin{figure*}
\begin{center}
  \subfloat{\includegraphics[width=0.9\textwidth]{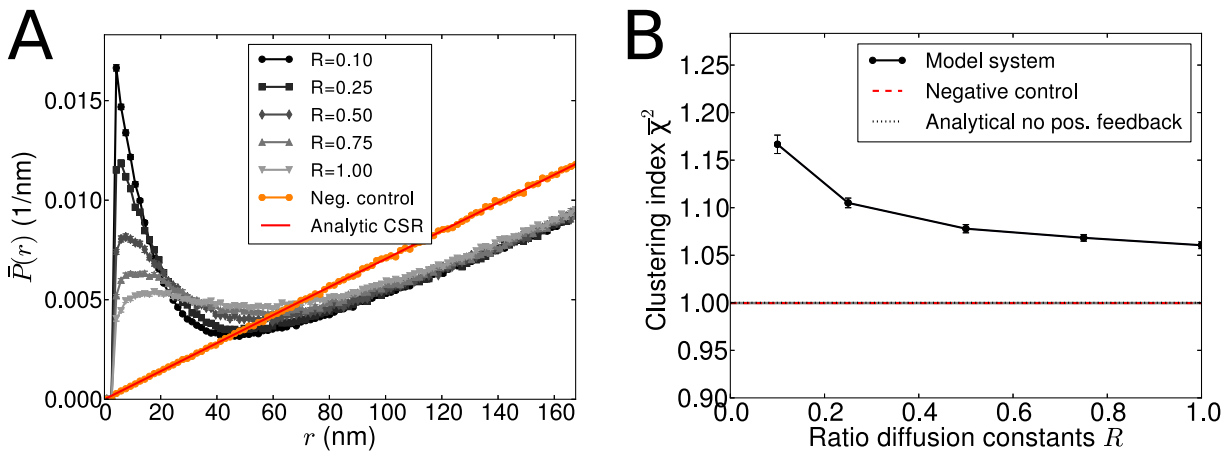}}  
\end{center} 
\caption{
Clustering persists with equal diffusion constants.
\textbf{A.} 
Distribution of interparticle distances $\bar{P}(r)$ for various values of the ratio $R$ of diffusion constants of the active to the inactive species.  Despite the fact that the extent of clustering, as indicated by the strength of the peak at low distances, decreases as $R$ increases as expected, the distribution remains significantly different from random (CSR) even at $R=1$.
\textbf{B.} 
Clustering index $\chi^2$ as a function of $R$. Consistent with A, $\chi^2$ decreases with $R$, yet at $R=1$ remains significantly larger than the CSR expectation of $\chi^2=1$.
In A error bars are the standard error of the mean, in B error bars are the $99\%$ confidence intervals. In A, these are smaller than the data points and therefore barely visible.
}
\label{fig:chis}
\end{figure*}

In the classic Turing picture of macroscopic pattern formation, clustering requires two species with sufficiently different diffusion constants \cite{Murray2003a}.  This requirement can be understood intuitively in the context of our system in the following way.  The active species should diffuse slowly, since then the local effect of positive feedback will outweigh the smoothing effect of diffusion and create clusters of active molecules.  At the same time, the inactive species should diffuse quickly, since then a cluster that has been nucleated will have a steady supply of inactive molecules to activate, and the cluster will persist.  
This intuition is consistent with 
the observation that the $\chi^2$ is significantly higher than $1$ 
when the ratio of diffusion constants of active to inactive molecules is less than one ($R=0.1$).
We now seek to determine for what range of diffusion ratios we continue to observe significant clustering.

Surprisingly, we find that significant clustering persists as $R$ is increased even to $R=1$, which corresponds to both the active and the inactive species having equal diffusion constants.  This is clear from Fig.\ \ref{fig:chis}, which shows both the distribution of interparticle distances $\bar{P}(r)$ and the clustering index $\chi^2$ for a range of diffusion ratios $R$.  As seen in Fig.\ \ref{fig:chis}A, as $R$ increases toward $1$, the amount of clustering decreases, as indicated by the reduction in the height of the peak at small distances.  Nonetheless, even at $R=1$, the distribution remains peaked and clearly different from the CSR expectation.  Correspondingly, as seen in Fig.\ \ref{fig:chis}B, the clustering index decreases as a function of $R$; yet, even at $R=1$, the clustering index $\chi^2 = 1.061 \pm 0.001 $ ($99\%$ confidence interval) is still significantly larger than the CSR expectation of $\chi^2 = 1$.

The fact that significant clustering persists even with equal diffusion constants underscores the stochastic nature of the clustering mechanism.  The number of molecules in each cluster is small---typically just a few---which is consistent with experimental observations (see Fig.\ \ref{fig:Hr}A in this manuscript and Fig. 1D in Eisenberg et al.).
Since the number is so small, and since activation is an inherently random reaction process, intrinsic fluctuations in this number are on the order of the number itself.  In such a case where stochastic effects dominate, it is not guaranteed that intuition drawn from a macroscopic analysis will carry over.
Indeed, in the context of yeast polarization, it has been observed in a model with one diffusing membrane species that stochastic effects lead to spatial heterogeneity even through a macroscopic analysis predicts a homogeneous solution \cite{Altschuler2008}.

In the next section, we analyze the macroscopic description of our model in detail, to understand more quantitatively the discrepancy between the intuition provided by studies of Turing patterns, and the observation that here clustering persists even with equal diffusion constants.

\subsection{The model lacks a Turing regime}

We now investigate quantitatively if and when clustering would be predicted in a macroscopic description of our system.  To this end we apply a standard technique for investigating Turing patterning \cite{Murray2003a}.
This technique assesses whether, when a spatially uniform distribution of molecules is perturbed slightly, it returns to the uniform distribution or it becomes more non-uniform.  In the first case, the uniform distribution is referred to as stable, whereas in the second case it is unstable.
Instability is then associated with pattern formation, since a small perturbation will drive the system away from uniformity and into a spatially heterogeneous stationary state.  The parameter regime leading to instability is then termed the Turing regime (see Appendix \ref{app:linstab} for more details).

The macroscopic description of our system is given by the deterministic rate equations that follow from Eq.\ \ref{eq:eqreaction},
\begin{eqnarray}
\label{eq:ode1}
\frac{dD}{dt} &=& \kappa_D\nabla^2 D-k_1D+k_2T-k_3DT,\\
\label{eq:ode2}
\frac{dT}{dt} &=& \kappa_T\nabla^2 T+k_1D-k_2T+k_3DT,
\end{eqnarray}
where $D(\vec{x},t)$ and $T(\vec{x},t)$ are the continuous concentrations of inactive and active molecules, respectively, and $\kappa_D$ and $\kappa_T$ are their respective diffusion constants.  In Appendix \ref{app:linstab}, we calculate the stability of the uniform stationary state by taking the Fourier transform and performing a linear stability analysis.  We find that the uniform stationary state is stable at all parameter settings.  In other words, there is no Turing regime.

This finding is perhaps not so surprising, given the simplicity of the model.  Inspecting Eqs.\ \ref{eq:ode1} and \ref{eq:ode2}, it is clear that the only nonlinearity is the $DT$ term that arises from the bimolecular reaction.  Typically, more complex nonlinear terms are required in order to support a Turing regime \cite{Murray2003a}.

On the other hand, this finding makes it all the more surprising that we observe clustering in our system at all, even when the diffusion coefficients are unequal.  Indeed, this finding strengthens the interpretation of the clustering we observe as an entirely stochastic effect, since clustering can never be possible according to the macroscopic model.  Even when the diffusion constants are unequal, which is where the extent of clustering that we observe is most significant (Fig.\ \ref{fig:chis}), we conclude that the origin of the clustering is 
purely due to the discreteness of the particle system.

Interestingly, previous work in the context of pattern formation has shown that intrinsic fluctuations can significantly extend the size of a Turing regime predicted by a macroscopic model \cite{Butler2009, Butler2011}.  Our finding supports this work, in the sense that intrinsic fluctuations lead to clustering at parameter settings for which no clustering is predicted macroscopically.  Our finding also extends this work, showing that clustering can emerge from intrinsic fluctuations in a model which supports no Turing regime at all.

\section{Discussion}

We have shown that positive feedback is sufficient to produce persistent, stochastic, dynamic clustering on cell membranes with statistics comparable to those of Ras signaling molecules.  
The parameter $\chi^2$ was used as an order parameter for the deviation from randomness,
and the $\chi^2$ values measured in our simulations are consistent with those calculated from experiments. 
Moreover, we have shown that the mechanism responsible for the clustering is purely stochastic, in the sense that a macroscopic, deterministic model of the system shows no Turing-like dynamic instability for any parameter settings.  In fact, we find that significant clustering persists even when the active and inactive species have equal diffusion constants, which goes against the intuition generally associated with Turing pattern formation.

We have focused on positive feedback because there is strong evidence in the Ras system for SOS-mediated positive feedback through allosteric binding \cite{Sondermann2004, Freedman2006}.  It is almost certain that this positive feedback is not the only factor contributing to Ras clustering.  Evidence for the involvement of the actin cytoskeleton comes from the observation that clustering is reduced upon the addition of an actin depolymerizing agent \cite{Plowman2005}, as well as from single-particle tracking studies showing transient compartmentalization of membrane molecules that is also modulated by cytoskeletal perturbations \cite{Kusumi2005}.  Evidence for the involvement of membrane domains comes from the observation that clustering is reduced upon cholesterol depletion \cite{Prior2003}.  Evidence for the involvement of complex formation comes from single-particle tracking studies showing the association of Ras slowdown with the binding of other signaling components \cite{Murakoshi2004} and from 
simulations and experiments suggesting that membrane-bound Ras forms dimers \cite{Guldenhaupt2012}.  Nonetheless, we find here that a minimal model of the positive feedback is sufficient to produce clustering statistics consistent with  experimental observations.  We therefore suggest that the mechanism identified here plays a role in seeding, maintaining, or reinforcing clustering that is also present due to these other factors.  
Indeed, such degeneracy of function is ubiquitous across many areas of biology, including biochemical signaling systems \cite{Edelman2001}.

In any system involving a species with two activation states, there are at least two distinct types of clustering: (i) all molecules can cluster, irrespective of their activation state,
or (ii) active molecules can cluster.
We focused on clustering of type ii because activation is explicitly linked to local density by the experimentally observed allosteric feedback.
Indeed, when Ras-GDP and Ras-GTP have the same diffusion constant, 
only clustering of type ii will be observed; clustering of type i will not be observed because
the total set of Ras particles (both D and T) will diffuse like indistinguishable hard spheres,
thus not showing clustering. 
However, 
when Ras-GDP and Ras-GTP have different diffusion constants, 
positive feedback combined with the difference in diffusion constants can lead to clustering of type i as well as type ii.
We emphasize that clustering of type i may also be caused by other mechanisms, e.g.\ interactions with membrane domains or the underlying cytoskeleton. In fact, experimental evidence exists for this possibility,
as it has been observed that Ras molecules containing only the membrane-binding domain, and not the nucleotide-binding domain that encodes the activation state, form clusters as well \cite{Prior2003, Plowman2005}.  Importantly, clustering of both types is likely to have similar consequences for signaling, since the downstream effector that propagates the signal would only respond to the active state, and in both cases the active 
molecules are viewed by this effector as clustered.
More experimental and theoretical study will be needed to determine whether a particular type of clustering dominates, and whether there is any associated effect on signal propagation.

We have focused on cases in which either (i) the activated species diffuses more slowly than the inactive species, or (ii) the two species have equal diffusion constants.  Single-particle tracking experiments have suggested that the diffusion of Ras molecules slows down upon activation \cite{Murakoshi2004}, lending support to the first case.  However, more recent single-particle tracking experiments have shown that both active and inactive Ras molecules exist in both a fast-diffusing and a slow-diffusing fraction \cite{Lommerse2005}.  Importantly, our results demonstrate that even if the diffusion of active and inactive molecules are comparable, significant clustering can emerge solely due to positive feedback and intrinsic noise.

We have refrained from extracting a specific cluster size, choosing instead to focus on the distribution of interparticle distances.  There are two key reasons for this choice.  First, the mechanism by which clusters arise in this study is a highly dynamic one: positive feedback competes with diffusion to support transient, locally dense activation events.  This mechanism results in clusters that are randomly seeded, have high turnover of member molecules, and, most importantly, have no well-defined boundary between ``cluster'' and ``non-cluster''.  Defining a cluster size, either in terms of a lengthscale or a typical number of molecules, is less natural in such a setting than in alternative settings, where clusters are generated by oligomerization or limited in number by binding to a scaffold protein.  Second, we prefer to make a comparison with experimental data at a stage which makes as few assumptions as possible.  Indeed, in experimental studies, it is common to further process primary interparticle-
distance data in the context of membrane domain models to extract a typical cluster size \cite{Prior2003, Plowman2005}.  Making a comparison with only the primary data reduces the number of assumptions we need to adopt beyond those which we clearly lay out herein in the context of our own model.

Stochastic heterogeneity is thought to play an important role in systems other than the Ras system.  For example, stochastic heterogeneity is thought to underlie oscillations and pattern formation in ecological predator-prey systems, beyond the predictions of macroscopic Turing models \cite{Butler2009, Butler2011}.  Stochastic heterogeneity---indeed driven by positive feedback---is also thought to play an important role in polarizing yeast cells prior to division \cite{Altschuler2008}.  Our study extends this field, showing that positive feedback is sufficient to produce stochastic clusters consistent with observed statistics of the Ras system.  At the same time, it is known that clusters at the membrane, both static \cite{Mugler2012} and cytoskeleton-induced \cite{Mugler2013}, are important for cell signaling properties.  It will be interesting to study the effects of positive-feedback induced clustering on the properties of Ras signaling, and the associated consequences for cell behavior.

\begin{acknowledgments}
The authors would like to thank 
Joris Paijmans, Thomas E.\ Ouldridge, and Nils B.\ Becker for helpful discussions, and
Joris Paijmans also for feedback on the manuscript.
This work is part of the research programme of the Foundation for Fundamental Research on Matter (FOM), which is part of the Netherlands Organisation for Scientific Research (NWO).
\end{acknowledgments}

\appendix

\section{Averaging procedures}
\label{app:averaging}

\subsection{Averaging $P(r)$, $H(r)$}

As described in the methods section, we run multiple simulations with the same
parameter settings. 
At fixed points in simulated time, the configuration of particles is 
acquired in each of these simulations. 
Let's for clarity explicitly define ${P}(r_j)\Delta r \equiv {P_s}(r_j,t_u)\Delta r$ and 
${H}(r_j)\Delta r \equiv {H_S}(r_j,t_u)\Delta r$ to be 
values obtained from a configuration in simulation $s$ at point $t_i$ in simulated time.
%
Averages $\bar{P}(r_j)\Delta r$ and $\bar{H}(r_j)\Delta r$ for each bin $j$ are determined by first determining
average $\bar{P}_s(r_j)\Delta r$ and $\bar{H}_s(r_j)\Delta r$ for each simulation separately,
averaging over the time points $t_i$ that are deemed to be in steady state.
%
%
Consecutively, to finally obtain $\bar{P}(r_j)\Delta r$ and $\bar{H}(r_j)\Delta r$ we average over 
the values for each simulation.
Also, we determine the standard error of the mean from the average values determined for each simulation.
Thus, mathematically, the averages $\bar{P}(r_j)\Delta r$ and $\bar{H}(r_j)\Delta r$ are obtained as follows:
\begin{eqnarray}
 \bar{P}(r_j)\Delta r \equiv \frac {1}{N_s N_t}  \sum_{s=0}^{N_s} \sum_{u=u_\text{s.s.}}^{N_t} {P_s}(r_j,t_u)\Delta r \nonumber \\
 \bar{H}(r_j)\Delta r \equiv \frac {1}{N_s N_t}  \sum_{s=0}^{N_s} \sum_{u=u_\text{s.s.}}^{N_t} {H_s}(r_j,t_u)\Delta r
\end{eqnarray}
Where $N_s$ is the total number of simulations performed, $u_\text{s.s.}$ is the index of time point $t_{u_\text{s.s.}}$ after which steady state is assumed, and $N_t$ is the total number of time points taken into account.
One can define:
\begin{eqnarray}
 \bar{P}_s(r_j)\Delta r \equiv \frac{1}{N_t} \sum_{u=u_\text{s.s.}}^{N_t} {P_s}(r_j,t_u) \Delta r\nonumber \\
 \bar{H}_s(r_j)\Delta r \equiv \frac{1}{N_t} \sum_{u=u_\text{s.s.}}^{N_t} {H_s}(r_j,t_u) \Delta r
\end{eqnarray} 
as intermediate averages over time points for each simulation $s$.
Using these intermediate averages standard error of the mean (SEM) values are determined for each bin $j$:
\begin{eqnarray}
 \text{SEM}_{P_j} \equiv 
    \frac {1}{\sqrt{N_s} N_s}  \sum_{s=0}^{N_s} (\bar{P}_s(r_j)\Delta r -\bar{P}(r_j)\Delta r)^2 \nonumber \\
 \text{SEM}_{H_j} \equiv 
    \frac {1}{\sqrt{N_s}N_s}  \sum_{s=0}^{N_s} (\bar{H}_s(r_j)\Delta r-\bar{H}(r_j)\Delta r)^2
\end{eqnarray}
Where the subscripts $P_j$ and $H_j$ are abbreviations for $\bar{P}(r_j)\Delta r$ and $\bar{H}(r_j)\Delta r$, respectively.


\subsection{Averaging $\chi^2$}
$\chi^2$ values are calculated for each configuration separately. 
As such, analogues to the previous section, values defined by 
equations \ref{eq:sigmasquared}, \ref{eq:sigmaCSR} and \ref{eq:chi2}
can be defined more sharply to be:
\begin{eqnarray} \label{appeq:sharpdef}
  \bar{\sigma}^2 		& \equiv & \bar{\sigma}^2_\text{s,u} \nonumber \\
  \sigma^2_\text{CSR}		& \equiv & \bar{\sigma}^2_\text{CSR,s,u}   \nonumber \\
  \chi^2			& \equiv &  \chi^2_\text{s,u}
\end{eqnarray}
Note that $\bar{\sigma}^2$ was already an average over particles $i$ (see equation \ref{eq:sigmasquared}). 
One could thus define:
\begin{eqnarray}
 \sigma^2_\text{s,u,i} \equiv
	  \frac{1}{M} \sum_{j=1}^{M} {[m_i(r_j)\Delta r - m_\text{CSR}(r_j)\Delta r]^2}
\end{eqnarray}
and
\begin{eqnarray}
 \bar{\sigma}^2_\text{s,u} \equiv
	  \frac{1}{M} \sum_{i=1}^{N_T} (\bar{\sigma}^2_\text{s,u,i})^2
\end{eqnarray}
Where $\bar{\sigma}^2_\text{s,u}$ is the quantity referred to as $\bar{\sigma}^2$ in Eq. \ref{eq:sigmasquared} (see also \ref{appeq:sharpdef}).
Note that analogously $\bar{\sigma}^2_\text{CSR,s,u}$ is an average over $\bar{\sigma}^2_\text{CSR,s,u,i}$ values,
but $\sigma^2_\text{CSR,s,u,i}$ has the same value for all $i$ values, and thus
$\bar{\sigma}^2_\text{CSR,s,u} = \sigma^2_\text{CSR,s,u,i}$.
Unlike the previous section, further averaging is not performed on $\bar{\sigma}^2_\text{s,u}$ 
and $\bar{\sigma}^2_\text{CSR,s,u}$ values.
Instead, these values are used to calculate $\bar{\chi}^2_\text{s,u}$ (as defined in Eq. \ref{eq:chi2}).
Consecutively, $\bar{\chi}^2_\text{s,u}$ values are averaged:
\begin{eqnarray}
 \bar{\chi}^2_\text{s} \equiv
	  \frac{1}{M} \sum_{u=n_\text{s.s.}}^{N_t} (\bar{\chi}^2_\text{s,u})^2
\end{eqnarray}
Which is again an intermediate average for a simulation $s$.
From these intermediate values, analogues to the previous section, overall averages and SEM values are determined:
\begin{eqnarray}
 \bar{\chi}^2 \equiv
	  \frac{1}{N_s} \sum_{s=1}^{N_s} (\bar{\chi}^2_\text{s})^2
\end{eqnarray} 
\begin{eqnarray}
 \text{SEM}_{\chi^2} \equiv 
    \frac {1}{\sqrt{N_s} N_s}  \sum_{s=0}^{N_s} (\bar{\chi}^2_s -\bar{\chi}^2)^2 
\end{eqnarray}
$99\%$ confidence intervals for $\bar{\chi}^2$ are extrapolated from the $\text{SEM}_{\chi^2}$ values. 
%



\section{A macroscopic model of our system is linearly stable}
\label{app:linstab}


To investigate if our system allows for clustering in a deterministic manner, 
the system is represented by a macroscopic model
and analyzed by a linear stability analysis to see if non-uniformity is expected.
%
Non-uniformity allows for the possibility of clustering.
A macroscopic model ignores the discrete nature of particles, describing the concentration of particles as a continuum in space and time.
Such a model consists of 
a set of equations that describe the change in concentration
in the system at all positions over time.
These equations
may contain non-linear terms.
When using a linear stability analysis, 
the non-linear macroscopic mathematical description is approximated by a linear macroscopic mathematical description.
%
Using this linear description, it is more easy to investigate the behavior of the system.
The question is whether a state of uniform concentration will return to 
uniformity after small disturbances (stable behavior) or whether it tends
to deviate further from uniformity (unstable behavior).
%
Unstable behavior thus allows for the possibility of clustering. 
This procedure is also described by Murray \cite{Murray2003a}.
%

%
%
Systems that are stable in the absence of diffusion, but unstable in the presence of diffusion are said to exhibit a diffusion-driven instability, and are sometimes called Turing unstable \cite{Murray2003a}.
These systems thus show no concentration fluctuations over time when averaging out the space component (thus ignoring diffusion),
but 
do show fluctuations when space and diffusion are considered.

In this appendix, we show that our system is not a Turing unstable system,
and we also show that it is not unstable in general.
We furthermore show that instability is also not expected for different diffusion constants.
In other words, we show that using a linear stability analysis on a macroscopic description of our system,
we do not expect deviations from a uniform concentration distribution.
Hence, clustering would not be expected based upon this analysis.

\subsection{A macroscopic description of the system}
Our system, as defined by the reactions in Eq. \ref{eq:eqreaction}, can be described by a set of differential equations as follows:
\begin{eqnarray}
\label{eq:ODEs}
\dot{T} & = & f(D,T)+R\nabla ^2T \nonumber  \\	 
\dot{D} & = & g(D,T)+\nabla ^2D ,
\end{eqnarray}
with
\begin{eqnarray}
\label{eq:fg}
f(D,T) & = & D-K T+S D T \nonumber \\
g(D,T) & = & -f (D,T) = -D+K T-S D T .
\end{eqnarray}
In these formulas, $T \equiv T(\vec x, t)$ and $D \equiv D(\vec x, t)$ describe
the concentration of respectively T and D particles at a point $\vec x \equiv (x_1,x_2)$ at time $t$ 
in a dimensionless form. 
To achieve this the following definitions were used:
%
\begin{eqnarray}
K  & \equiv &  \left[ \frac{k_2}{k_1} \right] , \nonumber \\
S  & \equiv &  \left[ \frac{k_3 P_0}{k_1} \right] , \nonumber \\
R  & \equiv &  \left[ \frac{\kappa_T}{\kappa_D} \right] , \nonumber \\
t  & \equiv &  \tilde t k_1 , \nonumber \\
\ell  & \equiv &  \sqrt{ \frac{\kappa_D}{k_1}} , \nonumber \\
\vec {x}  & \equiv &  \left[ \frac{\vec {\tilde x} }{ \ell} \right] , \nonumber \\
\nabla^2  & \equiv &  \ell^2 {\tilde{\nabla}}^2 , \nonumber \\
{T}(\vec x, t)  & \equiv &  \frac{1}{P_0} \tilde {T} (\vec{x}, t) , \nonumber \\
{D}(\vec x, t)  & \equiv &  \frac{1}{P_0} \tilde {D} (\vec{x}, t) .
\end{eqnarray}
$k_1$, $k_2$ and $k_3$ are the rate constant as defined in Eq. \ref{eq:eqreaction}, and their units 
are respectively s$^{-1}$, s$^{-1}$ and m$^{2}$s$^{-1}$.
$K$ is the equilibrium constant of the left reaction defined by Eq. \ref{eq:eqreaction}.
$S$ is envisioned to set the ratio between the speeds of the equilibrium reaction (left in Eq. \ref{eq:eqreaction}) and the positive feedback reaction (right in Eq. \ref{eq:eqreaction}). 
$R$ is the ratio between the two diffusion constants of the D and T particles, $\kappa_D$ and $\kappa_T$ respectively. 
Both diffusion constants have units of m$^2$s$^{-1}$.
$P_0$ is the total concentration ($/$m$^{2}$) of T and D particles. 
$\tilde t$ is the time in s.
$\ell$ defines the length scale, using the diffusion constant of D particles, $\kappa_D$.
$\tilde {D} (\vec{x}, t)$ and $\tilde{D}(\vec x, t)$ both give the number of particles ($/$m$^{2}$).
For all parameters we are only interested in values greater than zero.

When ignoring the space component, the equilibrium concentrations for these equations can be obtained
by realizing that without space $T + D = 1$ and solving $\dot D = 0$ and $\dot T = 0$. 
The equilibrium values $D^*$ and $T^*$ for respectively the D and T particles are:
\begin{eqnarray}
\label{eq:TstarDstar}
T^*_{\pm} & = & \frac{-1-K+S\mp \sqrt{4 S+(S-1-K)^2}}{2 S}  ,\nonumber \\
D^*_{\pm} & = & \frac{1+K+S \pm \sqrt{K^2-2 K (-1+S)+(1+S)^2}}{2 S} . \nonumber \\
\end{eqnarray}
Simple plotting learns us that $T^*_{-}$ and $D^*_{-}$ have values between $0$ and $1$, 
and thus are the physical solutions. These will be referred to as simply $T^*$ and $D^*$ from
hereon.

\subsection{Linearizing the macroscopic model}
To linearize this model the equations (\ref{eq:ODEs}) are Taylor expanded around
the equilibrium values $T^*$ and $D^*$.
The substitutions
$T = T^* + \delta T$ and
$D = D^* + \delta D$ 
are used to achieve this.
$\delta T$ and $\delta D$ represent small deviations from the equilibrium values $T^*$ and $D^*$.
Additionally, it is convenient to Fourier transform the equations using
\begin{equation}
 \hat h (k) \equiv \int_{-\infty}^{\infty} h(x) e^ {- 2 \pi i k x}
, 
\end{equation}
with h an arbitrary function.
When second order Taylor terms are neglected, and one realizes that
derivatives of $T^*$ and $D^*$ are $0$ because these are constants,
this leads to the following equations:
\begin{eqnarray}
\delta \dot T & = & f^*_T \delta T + f^*_D \delta D - R k^2 \delta T , \nonumber \\
\delta \dot D & = & g^*_T \delta T + g^*_D \delta D - k^2 \delta D
.
\end{eqnarray}
The partial derivatives of $f$ and $g$ should be evaluated at equilibrium concentrations of D and T, 
which is indicated by an asterisk ($*$).
These equations concern Fourier space, and from hereon this will be the case for all equations.
These equations then describe how the concentrations deviate from the equilibrium
values $T^*$ and $D^*$.
It now becomes convenient to use a matrix notation to express these equations:
\begin{eqnarray}
\label{eq:linearODE}
 \left(
\begin{array}{c}
 \delta \dot T \\
 \delta \dot D \\
\end{array}
\right)
& = &
\bm{J}^*
\left(
\begin{array}{c}
 \delta T \\
 \delta D \\
\end{array}
\right)+\left(
\begin{array}{cc}
 -R k^2 & 0 \\
 0 & -k^2 \\
\end{array}
\right)\left(
\begin{array}{c}
 \delta T \\
 \delta D \\
\end{array}
\right)
\nonumber \\
& \equiv &
\bm{M} 
\left(
\begin{array}{c}
 \delta T \\
 \delta D \\
\end{array}
\right)
, 
\end{eqnarray}
with J the Jacobian matrix
\begin{equation}
\label{eq:Jacobian}
 \bm{J} \equiv \left(
\begin{array}{cc}
 f_T & f_D \\
 g_T & g_D \\
\end{array}
\right) 
,
\end{equation}
and the last line defining $\bm M$. 
Note that the first term on the right in the first line describes the chemical reactions
and the second term on the right on that line describes the diffusion of particles.
(Again, the asterisk - $*$ - indicates the derivatives should be evaluated at equilibrium
concentrations.)
The partial derivatives in equation (\ref{eq:Jacobian}) can be deduced
in a straightforward way from equations (\ref{eq:fg}). 
This leads to the following equations:
\begin{eqnarray}
\label{eq:partialderivs}
f^*_T & = & -K + S D^* \nonumber \\
f^*_D & = & 1 + S T^* \nonumber \\
g^*_T & = & -f^*_T = K - S D^* \nonumber \\
g^*_D & = & -f^*_D = - 1 - S T^*  
\end{eqnarray}
The $T^*_{-}$ and $D^*_{-}$ solutions are found in equations (\ref{eq:TstarDstar}).
The analysis of systems as described by equation (\ref{eq:linearODE}) is well described in literature \cite{Murray2003a, Strogatz1994}, 
and will be the topic of subsequent sections.

\subsection{The stability of a linear system can be determined from its matrix}
When considering $(T, D)$ space for a point in $\vec x$ space, 
the two equilibrium concentrations $T^*$ and $D^*$
define a fixed point in $(T, D)$ space.
If straight line trajectories originate from this fixed point, they look like \cite{Strogatz1994}:
\begin{equation}
 (\delta T (t), \delta D (t)) =c_1e^{\lambda_1 t}\vec{v}_1+c_2e^{\lambda_2 t}\vec{v}_2
\end{equation}
With $\vec v$ a vector in $(\delta T (t), \delta D (t))$ space, $c_1$ and $c_2$ constants to be determined
from initial conditions. Substituting this solution in equation (\ref{eq:linearODE}), 
it can be seen that $\lambda_1$ and $\lambda_2$ are eigenvalues of the eigenvectors
$\vec{v}_1$ and $\vec{v}_2$ in
\begin{equation}
\label{eq:strogatzeigen}
\bm{M}
\vec v
=
\lambda
\vec v
,
\end{equation}
where $\bm M$ was defined in equation (\ref{eq:linearODE}).
$\vec{v}_1$ and $\vec{v}_2$ thus give the direction of the straight line trajectories.
(See the book by Steven H. Strogatz for a more elaborate discussion \cite{Strogatz1994}.)
Furthermore, a fixed point is considered
stable when both $\lambda$ values are smaller than $0$. As equation (\ref{eq:strogatzeigen})
is an eigenvalue problem, the values of $\lambda$ can be determined by solving
the characteristic equation. 
The solutions are \cite{Strogatz1994}:
\begin{equation}
\label{eq:lambdas}
 \lambda_{\pm} = \frac{1}{2} (\tau \pm \sqrt{\tau^2 - 4 \Delta}),
\end{equation}
with $\tau$ the trace of matrix $\bm M$ and $\Delta$ the determinant of matrix $\bm M$:
\begin{eqnarray}
\label{eq:deltatau}
 \Delta & = & \text{Det} ( \bm M ) = \lambda_- \lambda_+, \nonumber \\
 \tau   & = & \text{Tr} ( \bm M ) = \lambda_- + \lambda_+. 
\end{eqnarray}

\subsection{The linearized macroscopic model is not a typical Turing system}
As mentioned, for the system to be called Turing unstable, 
it should be stable in the absence of diffusion.
In the absence of diffusion the terms describing diffusion can be ignored, and
$\bm M$ in equation (\ref{eq:linearODE}) simply becomes $\bm J^*$. 
For clarity, from hereon whenever referring to this more simple matrix,
symbols will be marked with an apostrophe ($'$).
In other words, equation (\ref{eq:linearODE}) reduces to:
\begin{eqnarray}
\label{eq:linearODEfork0}
 \left(
\begin{array}{c}
 \delta \dot T \\
 \delta \dot D \\
\end{array}
\right) & = &
\left(
\begin{array}{cc}
 f^*_T & f^*_D \\
 -f^*_T & -f^*_D \\
\end{array}
\right) 
\left(
\begin{array}{c}
 \delta T \\
 \delta D \\
\end{array}
\right)
\nonumber \\
& = &
\left(
\begin{array}{cc}
 -K + S D^* & 1 + S T^* \\
 K - S D^* & - 1 - S T^* \\
\end{array}
\right) 
\left(
\begin{array}{c}
 \delta T \\
 \delta D \\
\end{array}
\right)
.
\nonumber \\
\end{eqnarray}
Note that the second line is identical to the first line times $-1$ because
$f = -g$ as described by equation (\ref{eq:fg}).
In general, when 
\begin{eqnarray}
 \Delta & > & 0,  	\label{eq:Turing1} \\
 \tau 	& < & 0,	\label{eq:Turing2}
\end{eqnarray}
equation (\ref{eq:deltatau}) shows both $\lambda'$ values
are ensured to be negative, and the solution to be stable.
Equations (\ref{eq:Turing1}) and (\ref{eq:Turing2}) are therefore two of four requirements
for a system to be classified as a Turing unstable (equations (\ref{eq:Turing3}) and 
(\ref{eq:Turing4}) discussed below give the two other requirements).
For our system, it follows directly using equation (\ref{eq:linearODEfork0}), 
\begin{equation}
\label{eq:ourdelta}
 \Delta' = \text{Det} (\bm M') = 0 ,
\end{equation}
and using equations (\ref{eq:linearODEfork0}) and (\ref{eq:TstarDstar}),
\begin{eqnarray}
\label{eq:ourtau}
 \tau' & = & f^*_T-f^*_D \nonumber \\
      & = & -\sqrt{(1+K-S)^2+4 S} < 0.	
\end{eqnarray}
This shows that when considering real values, 
only the second condition for the system to be Turing unstable, equation (\ref{eq:Turing2}), is met.
As equation (\ref{eq:lambdas}) shows, the fact that $\Delta' = 0$ results in eigenvalues of
\begin{eqnarray}
\label{eq:ourlambdas}
{\lambda_\pm}' & = & \{0, \tau'\} \nonumber \\
	    & = & \{0, (f^*_T-f^*_D)\} .
\end{eqnarray}
Using equations (\ref{eq:ourtau}) and (\ref{eq:ourlambdas}) is is clear that one $\lambda'$ value is negative,
and the second 0.
Combining equation (\ref{eq:strogatzeigen}) with equation (\ref{eq:linearODEfork0}), we deduce the 
respective eigenvectors are
\begin{eqnarray}
{\vec v}_{1}' & = & (1, - \frac{f^*_D}{f^*_T}) = (1, - \frac{1 + S T^*}{-K + S D^*}) , \nonumber \\
{\vec v}_{2}' & = & (1, -1).
\end{eqnarray}
Because we require the total amount of D and T particles to remain the same, 
we are only interested in behavior along the second eigenvector ${\vec v}_{2}'$.
The fact that we like to remain on this vector is already implicit in 
equations (\ref{eq:fg}), as $g$ is defined as $-f$.
Conversely, moving along the first eigenvector ${\vec v}_{1}'$ implies changing the amount
of particles in the system.

In any case, the values of $\lambda_\pm'$ being negative and zero shows that the system is expected to be 
stable when diffusion is not taken into account. However, because $\Delta'=0$ it can strictly not be 
classified as a Turing unstable system according to \cite{Murray2003a}.



\subsection{The linearized model is stable, not indicating clustering}
A more interesting question is perhaps whether the system including diffusion
is expected to be stable or unstable.
To analyze the system including diffusion, matrix $\bm M$ takes the more elaborate
form as described by equation (\ref{eq:linearODE}).
For the system to now be unstable, we need the opposite of equations (\ref{eq:Turing1})
and (\ref{eq:Turing2}) to be true for $\bm M$.
Thus, for instability we require one of either 
\begin{eqnarray}
 \Delta & < & 0,  	\label{eq:diffusionunstable1} \\
 \tau 	& > & 0,	\label{eq:diffusionunstable2}	
\end{eqnarray}
to be true (or both). Combining equation (\ref{eq:ourtau}) with equation (\ref{eq:linearODE}),
it can be seen that 
\begin{equation}
 \tau = \tau' - (1+R)k^2 < 0;
\end{equation}
this is smaller than zero because $R$ and $k$ are positive.
Instability thus cannot result from $\tau$ being bigger than zero (equation (\ref{eq:diffusionunstable2})).
Thus, only when equation (\ref{eq:diffusionunstable1}) holds, instability is expected
in the model also considering diffusion.
$\Delta$ can be calculated from equation (\ref{eq:linearODE}) to be the following:
\begin{equation}
 \Delta = k^4 R-k^2\left( f^*_T+ R g^*_D\right)+\Delta'
\end{equation}
(Note that $\Delta'$ was found to be zero in equation (\ref{eq:ourdelta}).)
From this equation, what is known as the third condition for Turing instability 
becomes clear: for $\Delta$ to be negative
\begin{equation}
 0 < ( f^*_T + R g^*_D ) \label{eq:Turing3}
\end{equation}
needs to hold \cite{Murray2003a}. Given equation (\ref{eq:ourtau}), for this to be true $f^*_T$
and $g^*_D$ need to be of opposite sign, and 
\begin{equation}
R \neq 1
\end{equation}
is a requirement. Using equations (\ref{eq:partialderivs}) as 
a reference, it is immediately clear that $g^*_D < 0$. For instability, we thus need
$f^*_T$ to be positive. 
However using equations (\ref{eq:TstarDstar}) and (\ref{eq:partialderivs})
\begin{eqnarray}
\label{eq:fTproof1}
 f^*_T & = & - \frac{1}{2} ( \sqrt{(1+K)^2-2 (-1+K) S+S^2} \nonumber \\*  
 & & - (S+1-K) ) < 0
\end{eqnarray}
since  
\begin{eqnarray}
\label{eq:fTproof2}
\sqrt{1+K^2+S^2+2S+2 K-2K S} \nonumber \\
> S+1-K \nonumber \\
\left< S+1-K > 0 \right> \nonumber \\
\Leftrightarrow \nonumber \\
1+K^2+S^2+2S+2 K-2K S \nonumber \\
> 1-2 K+K^2+2 S-2 K S+S^2 \nonumber \\
\Leftrightarrow \nonumber \\
2 K>-2 K.
\end{eqnarray}
holds ($S+1-K > 0$ is required for the sign not to flip between the first and second step) and $f^*_T$ is straightforwardly negative in Eq. \ref{eq:fTproof1} when $S+1-K < 0$.

It is thus clear that condition (\ref{eq:Turing3}) can not be met,
irrespective of the value of $R$. 
As this condition
was a requirement for both Turing instability and instability in general,
it is immediately clear that our system cannot be Turing unstable nor unstable in general.
As mentioned, based upon this analysis, that means clustering is not expected.
It is furthermore noteworthy that these conclusions are independent of the value of $R$.

%
For the sake of completeness, the fourth condition for a system to be 
categorized as Turing unstable is \cite{Murray2003a}
\begin{equation}
\label{eq:Turing4}
 (f^*_T+R g^*_D )^2-4 R (f^*_T g^*_D-f^*_D g^*_T) > 0.
\end{equation}
Because condition (\ref{eq:Turing3}) could not be met, this condition is 
irrelevant for our system.
\bibliography{./library}

%

\end{document}